\documentclass[aps,prb,reprint,superscriptaddress]{revtex4-1}

\usepackage{amsmath}
\usepackage{amsfonts}
\usepackage{subfig}
\usepackage{graphicx}
\usepackage{url}

\usepackage{dcolumn}% Align table columns on decimal point
\usepackage{bm}% bold math

\begin{document}

\title{Electron transport properties of sub-3-nm diameter copper nanowires}

\author{Sarah L.~T.~Jones}
\affiliation{Tyndall National Institute, University College Cork, Dyke Parade, Cork, Ireland}
\author{Alfonso Sanchez-Soares}
\affiliation{Tyndall National Institute, University College Cork, Dyke Parade, Cork, Ireland}
\author{John J.~Plombon}
\affiliation{Intel Corporation, Hillsboro, Oregon 97124, USA}
\author{Ananth P.~Kaushik}
\affiliation{Intel Corporation, Hillsboro, Oregon 97124, USA}
\author{Roger E.~Nagle}
\affiliation{Intel Ireland, Collinstown, Leixlip, Co. Kildare, Ireland}
\author{James S.~Clarke}
\affiliation{Intel Corporation, Hillsboro, Oregon 97124, USA}
\author{James C.~Greer}%
 \email{jim.greer@tyndall.ie}
\affiliation{Tyndall National Institute, University College Cork, Dyke Parade, Cork, Ireland}

\begin{abstract} 
Density functional theory and density functional tight-binding are applied 
to model electron transport in copper nanowires of approximately 1 nm and 
3 nm diameters with varying crystal orientation and surface termination. 
The copper nanowires studied are found to be metallic irrespective of 
diameter, crystal orientation and/or surface termination. 
Electron transmission is highly dependent on crystal orientation and
surface termination. 
Nanowires oriented along the [110] crystallographic axis 
consistently exhibit the 
highest electron transmission while surface oxidized nanowires show 
significantly reduced electron transmission compared to unterminated nanowires. 
Transmission per unit area is calculated in each case, for a given crystal 
orientation we find that this value decreases with diameter for unterminated 
nanowires but is largely unaffected by diameter in surface oxidized nanowires for 
the size regime considered. 
Transmission pathway plots show that transmission is larger at the 
surface of unterminated nanowires than inside the nanowire and that 
transmission at the nanowire surface is significantly reduced 
by surface oxidation. 
Finally, we present a simple model which explains the transport per unit 
area dependence on diameter based on transmission pathways results. 
 
\end{abstract}

\maketitle

\section{\label{sec:Intro}Introduction}
Due to continued nanoelectronics scaling, metal interconnects as well as 
transistors in integrated circuits are becoming ever smaller and 
are approaching atomic  scale dimensions \cite{ITRS}. 
As a consequence, understanding the effects of size-dependent phenomena on 
material properties is becoming ever more critical to enable efficient device 
performance. 
Present day transistor technologies use copper as an interconnect material, 
however it is unknown if at the small cross-sections required for future 
technologies copper will be able to fulfill this function effectively due 
to increased resistances; measured line resistivity increases dramatically 
for Cu nanowires (NWs) compared to bulk materials.
\cite{Steinhogl2002,Zhang2007} 
It is required that the electrical conductivity of small cross section 
copper nanostructures be understood to maintain acceptable power consumption in 
future nanoelectronics generations. 
In particular the contribution of the individual scattering sources, such 
as surfaces,\cite{Fuchs1938} grain boundaries,\cite{Mayadas1970} electron-phonon interaction \cite{Plombon2006} and 
impurities, to overall resistivity needs to be assessed to aid development 
of interconnects which minimize line resistance.
A succinct overview of these issues can be found in the review of 
Josell, Brongersma and T\H{o}kei.\cite{Josell2009}

In general the dominant scattering mechanism in metal nanostructures is 
dependent on processing conditions and the geometry of the resulting 
nanostructures.
For example \citet{Henriquez2010,Henriquez2013} have reported electron scattering 
to vary with grain size in the case of gold thin films.
Grain boundary scattering is found to dominate for grain sizes much smaller 
than the electron mean free path, while for much larger grain sizes 
it is the surface that is critical to resistivity. 
For grain sizes comparable to the mean free path, both mechanisms are 
responsible for the increases resistances relative to bulk gold. 
Meanwhile, in the case of copper thin films \citet{Sun2009,Sun2010} have reported 
that it is grain boundary scattering which dominates.
\citet{Zhang2007} report that the temperature dependence 
of resistivity in Cu NWs is consistent with surface scattering 
as the dominant scattering source. 
This report is corroborated by \citet{Graham2010}, 
who find that diffuse surface scattering and line edge roughness is 
consistent with the temperature dependence of resistivity for Cu NWs down 
to 25 nm diameter and that the role of grain boundary scattering is minimal. 
Similarly, \citet{Wang2012} report that resistivity is dominated by 
diffuse surface scattering for 20 - 100 nm Cu NWs. 
Thus though the situation for metal nanostructures can vary by material 
and process conditions, the literature to date as pertains to Cu NWs 
seems quite emphatic; the surface is the dominant scattering source.

In low dimension Cu films the surface environment has been reported to 
drastically influence conductivity. 
In a series of studies,\cite{Chawla2009,Chawla2010,Chawla2011} Chawla and 
co-workers have investigated the scattering at the surfaces of Cu thin films. 
They report that scattering at a Cu-vacuum surface is partially specular, 
whereas after tantalum deposition the surface scattering becomes diffuse
.\cite{Chawla2009,Chawla2011} 
The effect of oxygen has also been reported, with diffuse scattering again 
reported after oxidation.\cite{Chawla2010} 
The effect of a variety of coating metals on Cu thin films has also been 
reported from first principles simulations.\cite{Zahid2010} 
Metals with a density of states (DOS) comparable to the DOS 
of surface Cu atoms (Al and Pd) lower 
resistivity, while those which do not (Ta, Ti and Ru) increase resistivity. 

Against this backdrop, it is probable that surface scattering is of 
particular importance for small diameter Cu NWs and thus we study 
unterminated (i.e., in vacuum) and oxidized 1 and 3 nm Cu NWs using 
\textit{ab initio} and semi-empirical computational 
techniques. 
A high degree of anisotropy of Cu NW conductivity with crystal orientation 
has previously been reported using tight-binding computational 
methods,\cite{Hegde2014} thus we also consider the effects of crystal 
orientation. 
We calculate transmission spectra and transmission pathways
\cite{Solomon2010} in each case for geometry optimized structures. 
We find a strong dependence of transmission on crystal orientation 
and termination. 
[110] NWs consistently show a larger transmission than [100] and [111] NWs. 
Surface oxidation reduces transmission substantially compared to 
unterminated NWs. 
Additionally we find that transmission even for 1 nm unterminated NWs is 
larger at the NW surface than the core due to the existence of surface states. 
We provide a simple analysis of transmission in nanoscale Cu 
wires, which accounts for the 
trends in transmission we obtain.

\section{\label{sec:Method}Method}

\begin{figure}[b]
\includegraphics[scale=1.0]{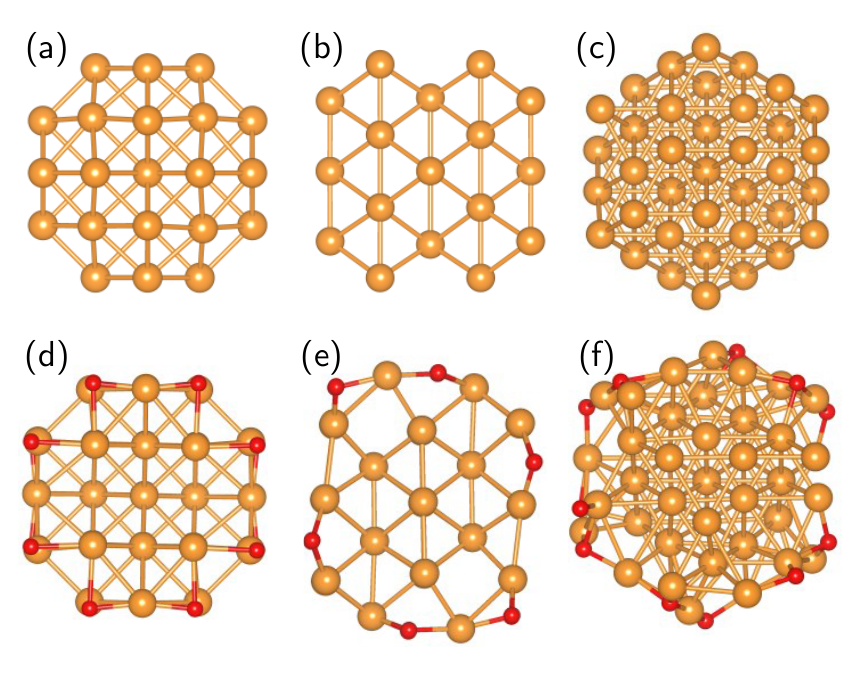} 
\caption{Structures of the geometry optimized 1 nm NWs. (a), (b) and (c) show 
the unterminated [100], [110] and [111] NWs, (d), (e) and (f) show the 
respective O-terminated NWs. Cu is orange and O is red.}
\label{fig:Small-Structures}
\end{figure}

\begin{figure}[h]
\includegraphics[scale=0.95]{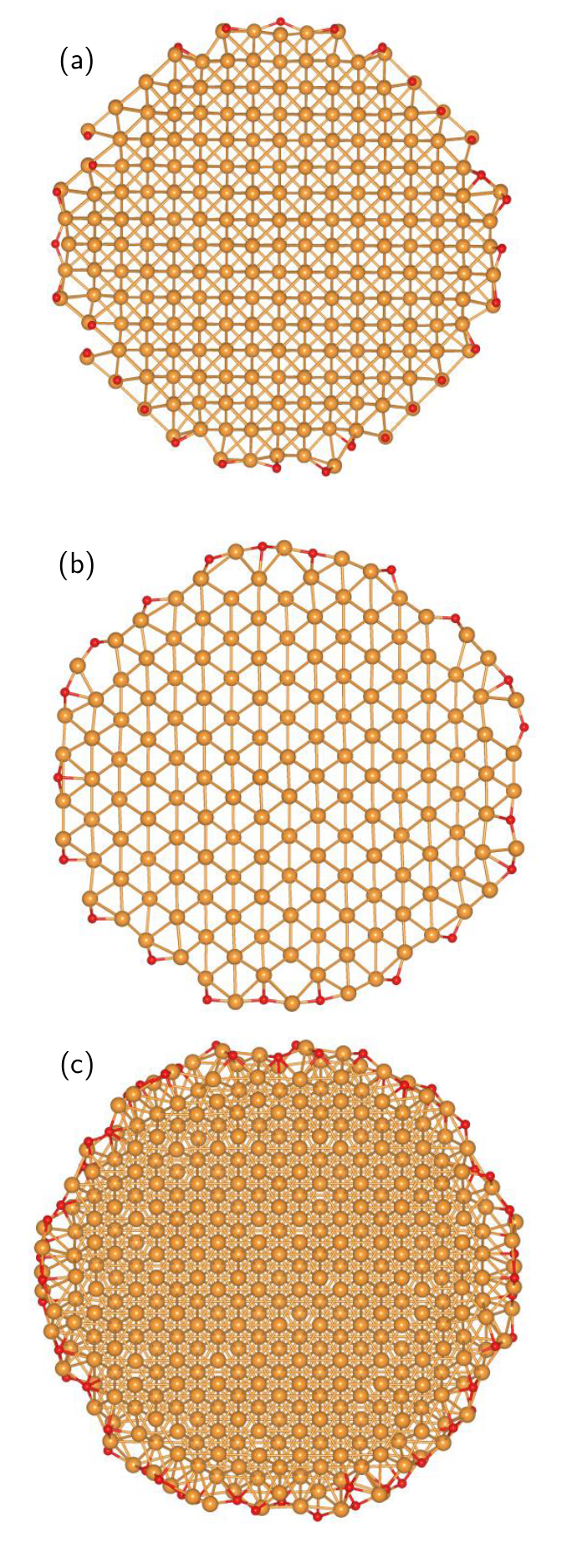} 
\caption{Structures of the geometry optimized O-terminated 3 nm NWs. (a), (b) 
and (c) show the [100], [110] and [111] NWs, respectively. Cu is orange and 
O is red.}
\label{fig:Large-Structures}
\end{figure}

The NW structures studied in this work are based on the bulk fcc Cu structure, 
with [100]-, [110]- and [111]-oriented NWs chosen with approximately 1 and 
3 nm diameters. 
Due to the small diameter of the NWs studied in this work, atoms at the NW 
surface will  have an especially large contribution to the NW properties 
e.g., 57\%, 59\% and 49\% of the atoms are at the surface for the [100], [110] 
and [111] 1 nm NWs, respectively. 
Hence in these nanostructures, electrical resistivity can be expected to 
be dominated by surface scattering. 
While standard complimentary metal oxide semiconductor (CMOS) 
technologies employ Ta based materials as barrier layers 
to prevent Cu diffusion, deposited Ta has been shown to cause diffusive 
scattering at the surface of thin Cu films.\cite{Chawla2011} 
As scaling continues, the relative contribution of surface scattering to 
total scattering increases dramatically and along with the barrier 
thickness means that Ta-based barrier layers may well become unsuitable 
for efficient interconnect design. 
To isolate barrier effects we have chosen to restrict this work to 
consideration of unterminated  (i.e., in vacuum) and oxygen terminated 
(i.e., surface oxidized) NWs. 
The relaxed (i.e., zero strain) structures of the NWs studied are shown in Fig.
\ref{fig:Small-Structures} and Fig. \ref{fig:Large-Structures}.

Density functional theory (DFT) as implemented in the OpenMX software package 
\cite{OpenMX} is used to study the electronic properties of the 1 nm diameter 
Cu NWs. 
The PBE\cite{PERDEW96} formulation of the generalized gradient approximation 
(GGA) exchange and correlation functional is used in conjunction with 
norm-conserving pseudopotentials\cite{MORRISON93} and a strictly localized 
pseudo-atomic orbital (PAO) basis,\cite{OZAKI03,OZAKI04} which enables 
us to perform the decomposition of the transmission into 
localized pathways.
The basis sets used are 6.0H-s4p2d2 and 7.0-s2p2 for Cu and O respectively. 
The first part of the basis set notation gives the PAO cutoff radius in Bohr, 
while the second part indicates the orbitals used for the valence electrons, 
e.g., O 7.0-s2p2 gives a cut-off radius of 7.0 Bohr and 8 basis functions 
(2 s functions and 6 p functions). 
This Cu basis set agrees reasonably with experiment, giving an optimal fcc 
lattice parameter of 3.63 \AA\ and a bulk modulus of 135 GPa compared to 
experimental values of 3.615 \AA\ and 137 GPa, respectively. 
The supercell approach is used and because the NWs are periodic only along 
their axis, a cell size incorporating a minimum of 1 nm of vacuum in the 
transverse directions is used such that the interaction between periodic 
images is minimized. 
The first Brillouin zone is sampled using 11 k-points generated according 
to the Monkhorst-Pack method \cite{Monkhorst1976} and  
an energy cutoff of 200 Ry is used to generate the grid by which real space quantities are discretized.
The atomic positions in the NWs are relaxed until all forces are 
less than 3x10$^{-4}$ Hartree/Bohr; total energy is also minimized with 
respect to the lattice cell parameter along the NW axis for each NW 
orientation. Optimized lattice parameters along the NW axis are given in 
Table \ref{tab:Cell_Parameter}. 

The 3 nm O-terminated NWs are also relaxed using the OpenMX 
software package using the same parameters. 
[100] and [110] NW structures are fully optimized whereas we take a 
hollow-core approach (described in the supporting information) for the [111] 
NW due to the number of atoms in the unit cell. Due to 
its reduced computational demand, Density Functional Tight Binding (DFTB) as implemented by the QuantumWise 
software package\cite{QW,Brandbyge2002,Soler2002} is used to calculate the electronic transport properties 
of the 3 nm NWs, using a density mesh cutoff of 15 Hartree 
and an 11 k-point Monkhorst-Pack grid. DFTB results were 
compared to DFT in a carefully chosen set of cases 
in order to explicitly assess their validity in NW structures.

The electrical properties of these 
NWs are modeled in the context of the Landauer-B\"{u}ttiker formalism, \cite{Landauer1957, Datta1995}
which includes the concepts of wide \emph{reflectionless} contacts and 
electrodes.  This formalism relates the electric current through 
a NW attached to two electron reservoirs (contacts) with the probability 
that an electron can be transmitted via the relation
\begin{eqnarray}
I = \frac{e}{h} \sum_{\sigma} \int T_{\sigma} (E,V)& \nonumber \\
\times \left[ f  \left( E , \mu_R , \right. \right. & \left. \left. T_R \right) -
f \left( E, \mu_L , T_L \right) \right] \, dE \, ,
\end{eqnarray}
where $e$ is the electron charge, $h$ Planck's constant, $T_{\sigma} (E,V)$ is
the transmission coefficient per spin channel $\sigma$ at energy $E$ and applied bias $V$, 
$f$ is the Fermi-Dirac distribution, $T_{L}$ ($T_{R}$) and $\mu_L$ 
($\mu_R$) are the temperature and chemical potential  
of the left (right) electrode, and the applied bias is given by

\begin{equation}
V=\frac{\mu_R - \mu_L}{e}.
\end{equation}
In the case of a NW directly attached to 
identical electron reservoirs (i.e., same chemical potential and temperature) 
we may write the linear response conductance in the zero-temperature limit as

\begin{equation}
G = \frac{I}{V}=\frac{e^2}{h} \sum_{\sigma} T_{\sigma}(E_F, V=0).
\end{equation}

Given that the NW is modeled as a
perfectly periodic structure, all Bloch waves propagate with unit probability. This leads to 
the expression for the NW contact resistance

\begin{eqnarray}
G_C^{-1} (E_F) =& \left[ \frac{e^2}{h} \sum_{i} T_i M_i \right]_{T_i = 1; E=E_F}^{-1} \nonumber \\
=& \left[ \frac{e^2}{h} M(E_F) \right]^{-1} \, ,
\end{eqnarray}

where the NW acts as a ballistic waveguides for the total number of modes at the Fermi level
$M(E_F)$, which includes all Bloch waves $\psi_{k \sigma}(E_F)$. This quantity 
provides information on the electrical resistivity intrinsic to the 
NW's electronic structure. Throughout this work the reported 
transmission then corresponds to a sum over the total 
number of modes $M(E)$ for each structure as a measure 
of their maximum potential conductivity as affected by 
their reduced scale and in the absence of 
scattering sources such as structural defects, 
grain boundaries, and phonons. Even though surface scattering is expected 
to be the most important contribution to resistance at these 
dimensions at zero temperature, the contribution of electron-phonon scattering 
at finite temperatures cannot be neglected.
\citet{Plombon2006} report a decomposition of the resistance 
of copper lines with widths ranging from 75 to 500 nm with electron-phonon 
scattering estimated to account for 60\% of the resistivity 
at 300K. The effects of electron-phonon interactions on 
resistivity in sub 5nm copper NWs are currently unknown and have been excluded from this study.

It is worth noting that for 
interconnect applications these NWs' function would be 
to provide electrical contact between devices (e.g., transistors)  across a short length and with small voltage drops 
- ideally zero. Hence the use of the linear response conductance 
applies and its validity can be extended to 
arbitrarily high temperatures as long as the transmission coefficient remains approximately constant over the energy 
range in which the transport takes place.

With the aim of providing some insight into 
the influence of the local chemical environment on 
electronic transport, 
a decomposition of transmission into spatially 
resolved pathways is computed according to the formalism 
described by \citet{Solomon2010}, as implemented 
in the QuantumWise software package, in which the transmission 
through a plane perpendicular to the transport direction 
that divides the system in two regions $A$ and $B$ is decomposed as

\begin{equation}
T(E) = \sum_{i \in A, j \in B} T_{ij}(E) \\,
\label{eq:pathways}
\end{equation}

where $i$ and $j$ are atoms on each side of said plane. 
We note that the use of a
Green's function implementation to the Landauer approach 
is required for decomposing the transmission into 
local contributions between pairs of atoms and hence this 
approach is taken for computing all electronic transport 
properties.

\section{\label{sec:Results}Results and Discussion}
\subsection{Nanowire structure}

The optimized cell parameters along the periodic direction of the NWs for 
a single repeat unit are given in Table \ref{tab:Cell_Parameter}. 
In the case of the unterminated NWs, the optimized cell parameter is very 
similar to the calculated bulk cell parameter;  
this agreement is reflected in the NW structures wherein 
the atoms move only slightly from their bulk positions after geometry 
optimization (see Fig. \ref{fig:Small-Structures}). 
When oxygen atoms are placed on the NW surface the situation changes 
dramatically with significant surface reconstruction seen for both 
the 1 nm and 3 nm NWs. 
There is a strong tendency for O to pull surface Cu atoms away from the 
NW bulk-like core until the bonded Cu and O atoms lie in a single plane. 
In bulk Cu(I) and Cu(II) oxide, each Cu bonds to 2 or 4 O atoms for Cu$_2$O 
and CuO respectively, conversely each O atom bonds to 4 or 2 Cu atoms for 
Cu$_2$O and CuO, respectively. 
The surface reconstruction seen in this work emulates the kinds of structural 
motifs seen in these oxides. 
The ratio of copper to oxygen (Cu:O) at the NW surface is 1.5, 1.7, and 1.9 
for the 1 nm oxidized [100], [110], and [111] NWs respectively, i.e., it lies 
between bulk CuO and bulk Cu$_2$O. 
In all cases, each O atom at the surface bonds to 3 Cu, thus the oxidation is 
intermediate to the situation seen for the bulk oxides. 
The surface oxide in the NWs consists of a very thin curved layer 
which locally is essentially two dimensional. 
Thus the NW surfaces are well oxidized as the Cu:O ratio is comparable 
to bulk copper oxide. 
For the [100] and [110] NWs, each Cu at the surface bonds to 2 oxygen atoms. 
Thus while the Cu:O ratio at the surfaces are different, in the case of [100] 
and [110] NWs, each O bonds to the same number of Cu atoms and each Cu bonds 
to the same number of O atoms. 
In spite of the similarity in the surface environment the surface geometries 
differ. 
For [100] the O-Cu-O motif is linear at Cu, while for [110] most O-Cu-O are 
V-shaped, a minority are linear. 
The [111] NW surface has 13 Cu atoms bonded each to 2 O atoms and 8 Cu atoms 
bonded each to 1 O atom, thus in line with the higher Cu:O ratio, this 
surface appears to be less oxidized than the [100] and [110] surfaces.
The Cu atoms at the [111] surface which bond to two O have a mix 
of linear and V-shaped O-Cu-O structural arrangements.

\begin{table}[h]
\caption{\label{tab:Cell_Parameter}Optimized cell parameter in \AA\ along the 
NW periodic direction. Unterminated NWs have cell parameter comparable to the 
calculated bulk value, while terminated NWs have a longer cell parameter.}
\begin{ruledtabular}
\begin{tabular}{l *{5}{c}}
& & \multicolumn{2}{c}{Unterminated} & \multicolumn{2}{c}{O-terminated} \\
& Bulk & 1 nm & 3 nm & 1nm & 3nm \\
\hline
{[}100{]} & 3.63 & 3.60 & 3.63 & 3.75 & 3.64 \\
{[}110{]} & 2.57 & 2.56 & 2.59 & 2.86 & 2.73 \\
{[}111{]} & 6.29 & 6.26 & - & 6.46 & -\\
\end{tabular}
\end{ruledtabular}
\end{table}

The surface reconstruction does not strongly affect the positions of Cu atoms 
below the surface layers, this is clearly evident in the 3 nm NW structures 
shown in Fig. \ref{fig:Large-Structures}, beyond the first two Cu surface 
layers the Cu atoms occupy approximately equivalent configurations with 
respect to their bulk positions. 
The reconstruction at the surface is associated with a lengthening of the 
optimized cell parameter relative to the bulk and unterminated NW, which 
is more severe for the 1 nm NWs than 3 nm NWs. 
This suggests that there is competition between the surface of the NW and 
core over the relaxed cell parameter along the axis.  
The oxidized surface favors an increase in length along this axis, 
presumably to better incorporate the surface O atoms in a layer. 
Conversely the core of the NW consists only of Cu atoms with their  
geometry similar to bulk fcc Cu and therefore favors a bulk-like cell 
parameter. 
Thus while unterminated NW cell lengths are similar for 1 and 3 nm NWs, the 
oxidized 3 nm NWs have a smaller (i.e., closer to the bulk value) cell 
parameter than the oxidized 1 nm NWs. 
A concomitant increase in NW diameter is not observed, therefore this change 
in cell parameter occurs exclusively along the NW axis. 
The surface structural rearrangement seen here for the O-terminated [100] NW, 
shown in Fig. \ref{fig:Small-Structures}(d) and Fig. 
\ref{fig:Large-Structures}(b), is very similar to that seen in experiment 
for surface oxidized Cu(100) slabs\cite{Baykara2013} 
suggesting the structural rearrangement we see for the NWs in our simulations 
is representative of experimental structures with highly prepared surface 
treatments.

\subsection{Electron transmission}
\begin{figure}
\includegraphics[scale=0.95]{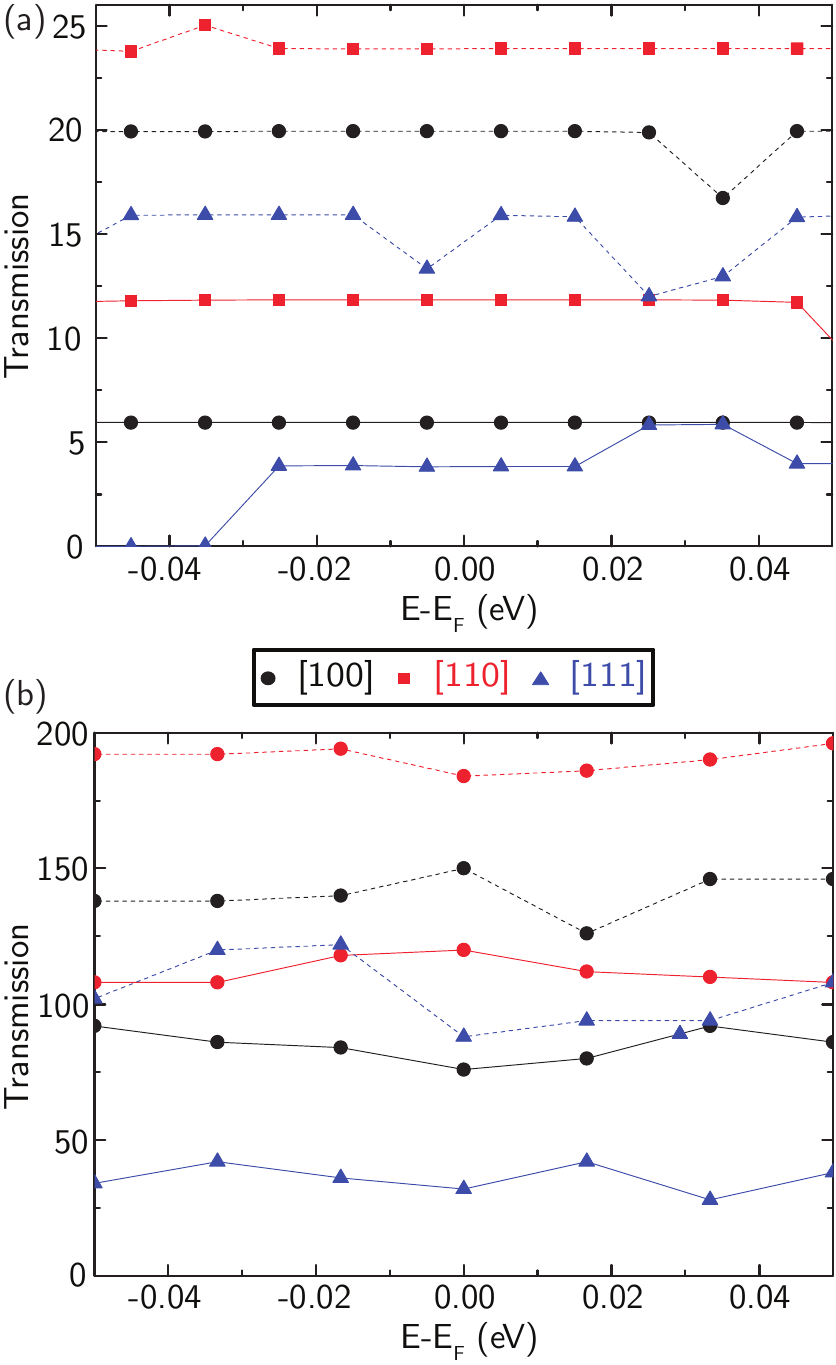}
\caption{Transmission for (a) 1 nm and (b) 3 nm Cu NWs. Unterminated NWs 
(dashed lines) have a larger transmission than oxidized NWs (full lines) 
irrespective of diameter or crystal orientation. Transmission is strongly 
influenced by crystal orientation, with [110] NWs having larger transmission 
than [100] and [111] NWs. The zero of energy is taken to be 
at the Fermi level.}
\label{fig:Transmission}
\end{figure}

\begin{table}[b]
\caption{\label{tab:Transmission_Area}Calculated transmission per unit area 
for unterminated and oxidized 1 nm and 3 nm NWs. The area of each NW is 
the area of a circle with diameter the sum of the NW width and twice the Cu atomic 
radius.}
\begin{ruledtabular}
\begin{tabular}{l *{5}{r}}
& & \multicolumn{2}{c}{Unterminated} & \multicolumn{2}{c}{O-terminated} \\
& Bulk & 1 nm & 3 nm & 1nm & 3nm \\
\hline
{[}100{]} & 28.0 & 22.8 & 19.4 & 7.3 & 8.8 \\
{[}110{]} & 29.7 & 26.7 & 24.3 & 15.3 & 15.1 \\
{[}111{]} & 27.9 & 15.3 & 11.8 & 3.7 & 4.0 \\
\end{tabular}
\end{ruledtabular}
\end{table}

The transmission spectra calculated for the terminated and oxidized 1 nm and 
3 nm Cu NWs are shown in Fig. \ref{fig:Transmission} for a narrow energy range 
of $\pm$50 meV centered about the Fermi energy.
In each case, we find that the oxidized NW has a substantially lower electron 
transmission than the corresponding unterminated NW. 
This result is consistent with the increase in resistance after chemical 
oxidation previously been reported by \citet{Chawla2010} 
for Cu thin films and arises from  the reduction of the DOS at 
the surface upon oxidation.
Additionally, we find a strong dependence of transmission on NW orientation. 
In all cases we find that the [110] NW has a larger transmission than [100], 
which in turn has a larger transmission than [111], irrespective of 
termination. 
As expected the transmission of the NWs increases going from 1 nm to 3 nm 
diameter due to the increasing cross-sectional area.
The cross-sectional areas of the NWs are not identical as diameters 
deviate from exactly 1 nm and 3 nm due to the different atomic arrangements of 
the crystal orientations. 
Thus, we also calculate the transmission per unit area $t$, presented in 
Table \ref{tab:Transmission_Area}, to ensure that this variation in 
transmission with orientation is not due to deviation in cross sectional 
area. 
There is ambiguity in how to define the cross sectional 
area of the NWs due to atomic structure. 
The atoms in the NW have volume which extends beyond the nuclear positions 
and the smaller the NW the greater this contribution will be to the NW 
cross section. 
For simplicity we have taken the cross-section to be circular,  elliptical, or octagonal 
as appropriate, with dimensions equal to the maximum inter-nuclear width of the 
NW plus twice the Cu atomic radius (1.28 \AA). 
We find that the transmission per area $t$ follows the same trend as the 
total transmission plotted in Fig. \ref{fig:Transmission}; [110] has 
the largest $t$ and [111] the smallest, regardless of diameter 
or termination.
This result is consistent with the anisotropy 
in Cu conductivity with crystal orientation reported recently by \citet{Hegde2014}.
In fact, this orientation dependence is of such significance that for the 
3 nm NWs the \textit{oxidized} [110] NW has a transmission comparable to  
the \textit{unterminated} [111] NW at the Fermi energy; 
a remarkable result consistent with the shape of the 
Fermi surface of bulk fcc copper, in 
which the Fermi surface exhibits a vanishing DOS along the  $\left\langle 111 \right\rangle$ directions. 
The extent of the orientation dependence on transmission is also clearly 
evident in Table \ref{tab:Transmission_Area}. 
In the case of unterminated NWs, the transmission for a given cross sectional 
area for [110] is approximately twice that of [111] NWs, after oxidation this 
margin further increases with the [110] transmission per unit area 
approximately 3.5 times larger than for the [111] NW. 
It may be anticipated that a more oxidized NW would show greater suppression 
in transmission, however the 1 nm [111] NW is in fact less oxidized than the 
[110] NW; each surface Cu in the [110] NW bonds to 2 oxygen atoms while 
surface [111] atoms bond to either 1 or 2 oxygen atoms. 
Meanwhile, unterminated [100] and [110] NWs show similar transmission 
per unit area; [100] is about 
4 units smaller than [110] for both 1 nm and 3 nm NWs. 
After oxidation, the transmission per unit area for [100] NWs is 
approximately 60\% of [110] NWs, i.e., transmission suppression on 
oxidation is relatively larger for [100] than [110]. 
Thus both the surface chemical environment and the NW orientation play 
a critical role in determining the overall transmission of the copper 
NW. 

\subsection{Local transmission paths}
\begin{figure}
\includegraphics[scale=1.0]{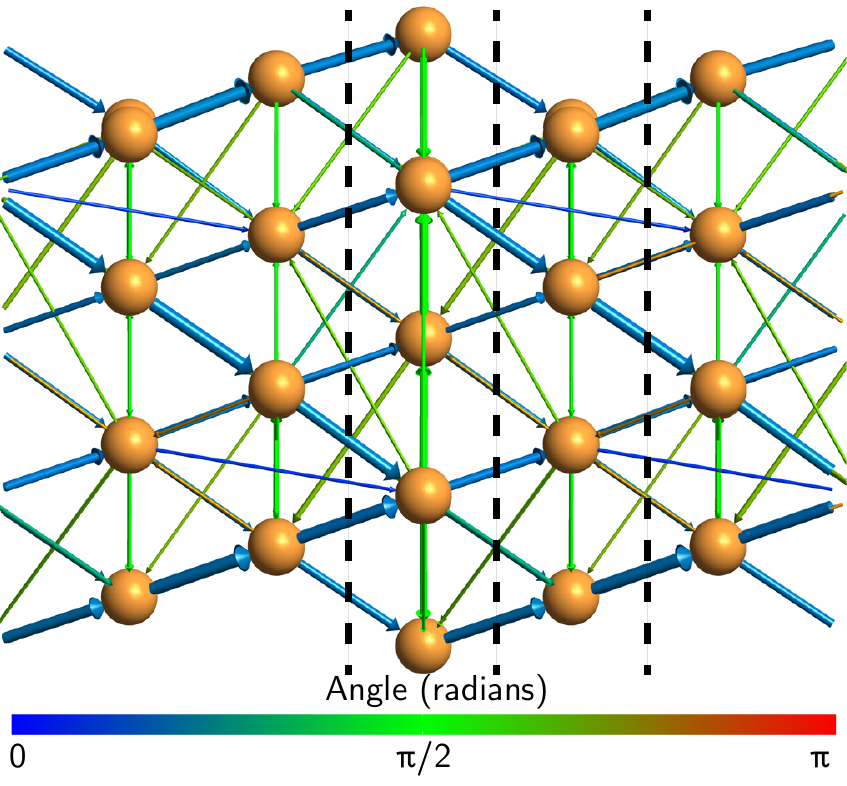}
\caption{Transmission pathways at the Fermi energy for 
unterminated 1 nm [111] Cu NW. 
The arrows show the direction of transmission paths. 
The arrow thickness represents the magnitude of the transmission 
while the color represents the orientation of the path 
with respect to the NW axis. Forward transmission is shown in 
blue, radial transmission in green, and backscattering in 
red. Pathways with transmission below 10\% of the maximum value 
are omitted for clarity. The dashed
black lines mark the positions of the planes shown Fig. \ref{fig:1nm-Cross-Section}(e).}
\label{fig:1nm-Pathways}
\end{figure} 

\begin{figure}
\includegraphics[scale=1.0]{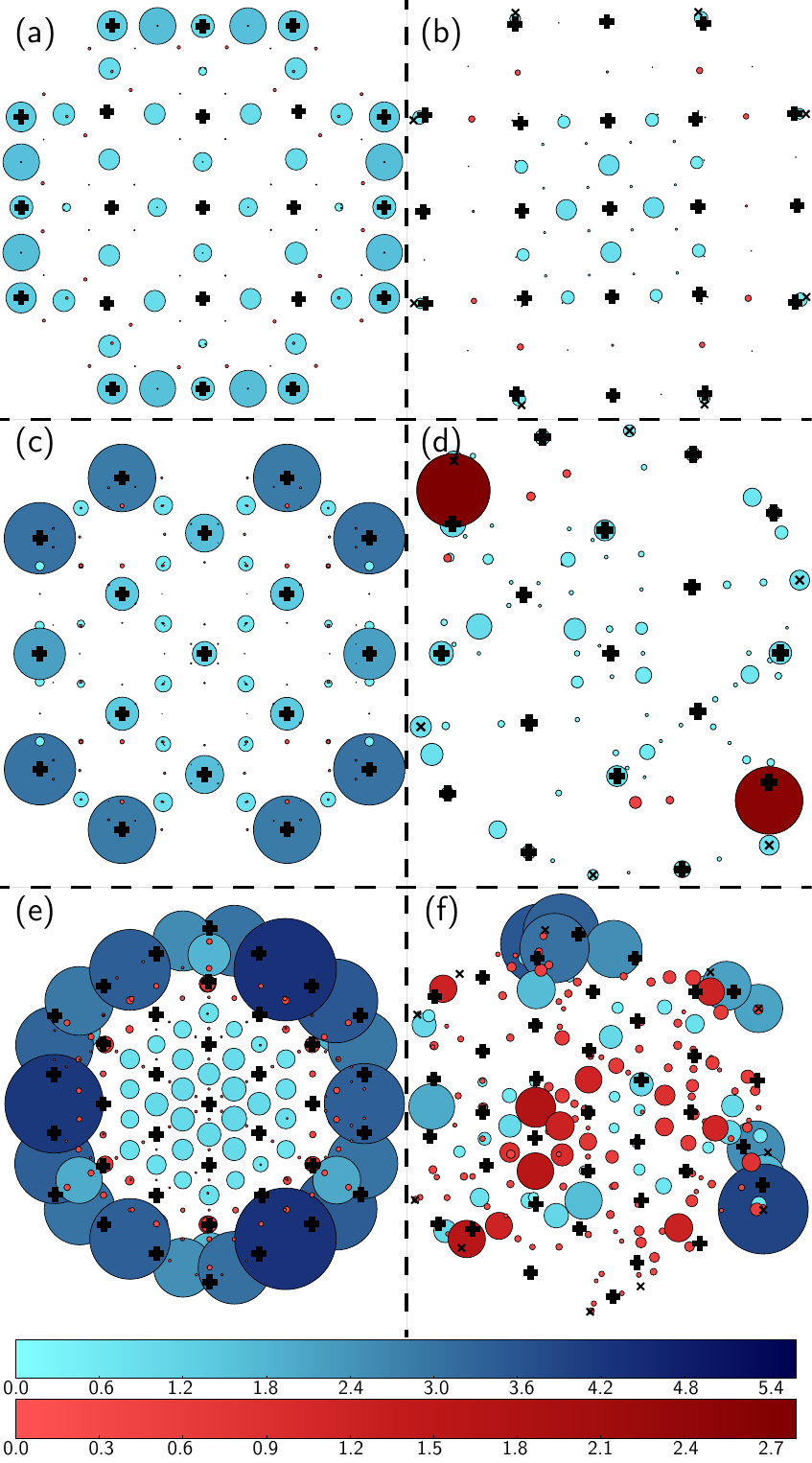}
\caption{Cross-section view of the transmission pathways at the 
Fermi energy. (a) Unterminated [100], 
(b) O-terminated [100], (c) Unterminated [110], (d) O-terminated [110], 
(e) Unterminated [111] and (f) O-terminated [111]. Atomic 
positions are shown as "+" (Cu) or "x" (O). The 'dot' size 
represents the absolute size of the transmission pathway crossing the plane, 
while the 'dot' color indicates the size and direction of the 
transmission pathway. 
Blue 'dots' correspond to forward transmission and red 'dots' to 
backscattering. Note the difference in the color scales ranges.}
\label{fig:1nm-Cross-Section}
\end{figure} 

The transmission results show that orientation and surface environment 
drastically impact on Cu NW electron transport. 
These effects are not easy to decouple, however, a greater 
understanding of the relationship between them can be established 
by examining transmission through the NW at an atomic scale. 
We have calculated transmission pathways in the NWs\cite{Solomon2010} at the Fermi level, which 
are shown in 
Fig. \ref{fig:1nm-Pathways} for the unterminated [111] NW, to particularly elucidate the effects of NW 
crystal orientation and surface oxidation on transmission.  
Transmission pathways provide a visual representation of transmission 
through the NW with the arrow width representing the size 
of the transmission between an atom pair.
As described in the caption of Fig. 
\ref{fig:1nm-Pathways}, the arrows are colored according to 
their angle with respect to the NW axis; blue arrows 
correspond to forward transmission, green arrows to radial 
transmission (i.e., atoms within the same transverse plane), 
and red arrows backscattering or reflection. As expressed in 
Eq. (\ref{eq:pathways}), the transmission pathways crossing a 
given plane perpendicular 
to the transport direction always sum to give the total 
transmission coefficient.

A more convenient visual representation is portrayed 
in Fig. \ref{fig:1nm-Cross-Section} in which 
the intersections of transmission pathways 
with planes located halfway between non-equivalent atomic planes 
give a clear image of the decomposition of the total transmission 
into pathways per cross-sectional area per NW. 
Such planes are indicated by 
dashed black lines in Fig. \ref{fig:1nm-Pathways} for 
the case of the unterminated [111] NW.
These projections provide 
a heuristic representation of localized transmission 
and allow quick
identification of where the most important contributions 
lie within the NW cross-section.

Fig. \ref{fig:1nm-Cross-Section}(a) shows transmission pathways for the 
unterminated [100] NW. 
There are 
transmission pathways both parallel ('dots' 
centered at atomic positions) and angled to the NW axis (centered 
between atomic positions). Angled pathways are larger 
in magnitude than parallel pathways as represented by their larger 
'dot' diameter and their color. 
This difference in magnitude can be 
attributed to the distance between the Cu atoms being larger if a direct route 
parallel to the axis is taken (3.60 \AA\ versus 2.53 \AA). 
It is is readily seen that pathways along the surface 
dominate electron transmission for this NW as the 'dots' in the 
surface are larger than below the surface, where the parallel
pathways are smaller and 
obscured by the symbols indicating the atomic positions.

The unterminated [110] NW transmission pathways are shown in Fig. 
\ref{fig:1nm-Cross-Section}(c). 
Transmission involving only surface atoms occurs along the NW axis and is  
large (indicated by the darker blue 'dots' centered about atomic 
positions) while pathways at an angle to the axis exhibit much lower transmission. 
Below the surface the situation is similar, transmission along the axis is larger than off-axis. 
In this orientation the Cu atom separation is similar for both parallel and 
off axis transmission and thus the more direct parallel route is favored. 
As for the unterminated [100] NW, transmission is larger at the NW surface 
than below the surface with the largest 'dots' corresponding to 
pathways located at the surface and parallel to the NW axis.
The transmission pathways for the unterminated [111] NW, shown in Fig. 
\ref{fig:1nm-Cross-Section}(e).
In this case, there is no significant transmission directly 
parallel to the NW axis, this is presumably 
because the Cu interatomic distance along the NW axis is simply too large 
compared to the off axis distances (6.26 \AA\ versus $\sim$2.5 \AA\ reflecting 
that no nearest neighbors align along the NW axis).
As previously mentioned, this result is consistent with the 
vanishing density of states found along the $\left\langle 111 \right\rangle$ directions in the Fermi surface of fcc Cu; there are no transmission paths parallel to 
the [111] direction as shown by the absence of arrows parallel to the 
NW axis in Fig. \ref{fig:1nm-Pathways} and of 
'dots' centered about atomic positions in Fig. \ref{fig:1nm-Cross-Section}(e).  
As shown in Fig. \ref{fig:1nm-Pathways} transmission pathways tend to flow along the NW surface until they reach  
a Cu atom with a low coordination number (such as the one at the 
top center of the figure), at which point the electrons 
are scattered into the NW core where they continue to propagate 
along angled paths until they reach another 
surface where this process repeats. The lack of transmission 
pathways parallel to the [111] NW axis makes the surface critical 
as all paths inevitably lead to the surface. 
While there are some transmission pathways inside the core of the 
NW, transmission 
\textit{through} the NW occurs primarily at the NW surface; and much like the 
unterminated [100] and [110] NWs, transmission is larger at the surface.
However, in this case as well as 'dots' representing forward transmission, 
the 'dots' representing negative values (i.e., electron 
backscattering or reflection paths) are much larger than for the 
unterminated [100] and [110] NWs.

\begin{figure}
\includegraphics[scale=1.0]{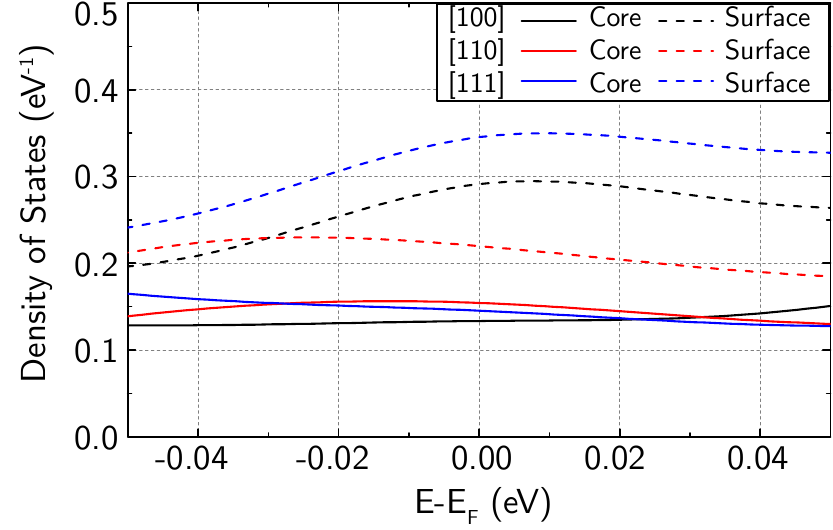}
\caption{Average local density of states for an atom at the surface 
(dashed) and at the center (solid) of unterminated 1nm NWs. The 
zero of energy is taken to be at the Fermi level.}
\label{fig:1nm-LDOS}
\end{figure}

The transmission pathways of the unterminated NWs consistently show 
that transmission is larger at the NW surface than in the core and also 
that backscattering for these NWs is relatively small, particularly 
for the [100] and [110] NWs in which the total magnitude of 
all reflecting paths is 14\% and 13\% of that of forward paths, 
whereas for [111] NWs this value increases to 
25\%. This surface-dominated 
transmission can be explained in terms of the increased local 
density of states (LDOS)
of surface atoms at the Fermi level as a result of their low 
coordination. This allows electrons otherwise localized 
at bonds to become available 
for transmission at the surface atoms. 
Fig. \ref{fig:1nm-LDOS} shows the average local density of 
states (LDOS) of an atom situated at the surface and of an 
atom close to the center 
of the NWs' cross section; the atoms at the surface show 
a larger density of states in the range around the Fermi level 
studied in this work. Unterminated [111] 
NWs exhibit the 
largest boost with the average LDOS value of a surface atom being 
2.55 times that of an atom near the center of the NW 
at the Fermi level; 
similarly, atoms at the surface of [100] NWs have an average 
surface-to-center LDOS ratio of 2.41 and a considerably 
lower ratio of 1.39 is found for [110] NWs.

As might be expected from the transmission spectra shown in  
Fig. \ref{fig:Transmission} and the previous analysis, oxidation 
of the NWs changes the behavior at the 
NW surfaces considerably. 
In the [100] NW, transmission at and just below the NW surface is 
reduced upon oxidation, in fact Fig. \ref{fig:1nm-Cross-Section}(b) 
shows that near the surface forward transmission 
is largely 
suppressed and backscattering pathways are enhanced.
In the core of the NW the transmission is similar to the 
unterminated NW as shown by similarly arranged, 
sized and colored 'dots'.
Thus in the case of the [100] NW, oxidation appears to affect the surface 
and immediate subsurface almost exclusively. 
In the case of the [110] NW, more significant back scattering is seen upon 
surface oxidation with the appearance of the large red 'dots' indicating 
a relatively large magnitude in Fig. \ref{fig:1nm-Cross-Section}(d).
Unlike the [100] NW, transmission is also reduced in the core of 
the [110] NW upon surface oxidation. 
A suppression of the pathways parallel to the NW axis can be 
clearly seen all the way to the center when comparing to the 
unterminated NW (\ref{fig:1nm-Cross-Section})(c).
Thus it appears that the effect of surface oxidation is felt deeper in the NW 
for the [110] orientation compared to [100]. 
However, transmission per unit area for the oxidized [100] NW actually drops 
more relative to the unterminated NW compared to the [110] orientation, as 
reported in Table \ref{tab:Transmission_Area}. 
This is likely caused by a reduction of transmission at the 
surface and immediate subsurface in the [100] oxidized NW, 
in which the surface has a larger
contribution relative to the core in the 
unterminated case as shown in Fig. \ref{fig:1nm-LDOS} and 
discussed above.
The surface of the oxidized [111] NW is more disordered than the oxidized 
[100] and [110] NWs studied in this work. 
This disorder along with the indirect transmission associated with transport 
along this crystal orientation leads to a complicated, disordered 
transmission pathways (Fig. \ref{fig:1nm-Pathways}(f)) consistent with the low 
transmission calculated for this structure. 
In contrast to the [100] and [110] oxidized NWs, there is large forward 
transmission at the surface for the oxidized [111] NW, which can be seen 
clearly in the mapping of the transmission pathways onto a cross sectional 
area of the [111] NW shown in Fig. \ref{fig:1nm-Cross-Section}(f). However, the large backscattering paths 
in the core of the NW result in a lower overall transmission; 
all paths in the core lead either to the surface (where 
oxidation has largely suppressed forward transmission) or are backscattered.

\subsection{Transmission model}

To understand the behavior of the total transmission per NW 
given in the previous sections, we provide an analysis for a simple 
model describing the electron transmission behavior of round NWs 
which partitions the NWs into regions of similar behavior 
based on the magnitudes of the transmission pathways within the 
them.
The generalization of the model to other geometries is straightforward. 
The model provides insight into the observation that transmission per 
unit area decreases as area increases for the unterminated NWs and that the 
transmission per unit area is relatively constant for oxidized NWs for a 
given orientation.

\begin{figure}
\includegraphics[scale=0.7]{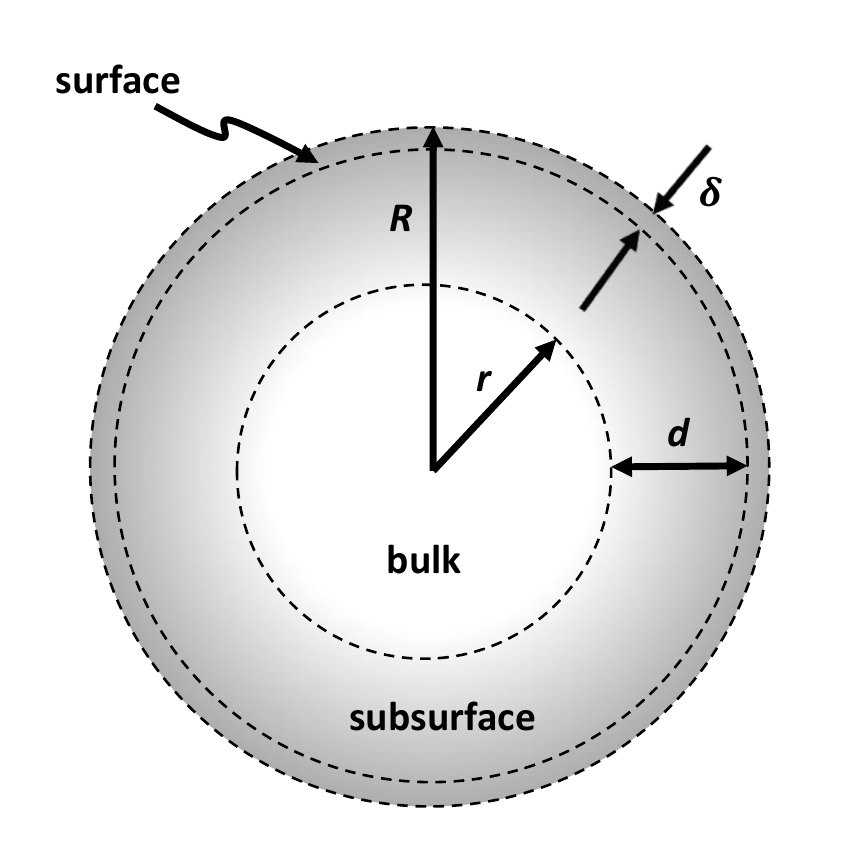}
\caption{Schematic representing the copper NW model, with surface, 
sub surface and bulk regions indicated. 
$R$ is the radius of the NW, $r$ is the radius of the bulk region,  
$d$ is the width of the subsurface region, i.e., the depth to which $t$ is 
affected by the surface, and $\delta$ is the width of the surface.}
\label{fig:Model}
\end{figure} 

The total transmission $T$ of a round NW of radius $R$ is divided into 
contributions from three distinct regions of the NW; the surface, the 
subsurface and the bulk regions; as shown in Fig. \ref{fig:Model} for a 
cross section of the NW. 
The surface region is described by a thickness $\delta$, the subsurface 
region by a thickness $d$ and the bulk-like core is described by a 
radius $r$; therefore

\begin{equation}
\label{eq:radius}
R = r + d + \delta.
\end{equation}

To each region is assigned a transmission per unit area $t$. 
The surface transmission per unit area $t_{surf}$ is assumed constant, 
whereas the subsurface transmission per unit area $t_{sub}$ can vary 
with depth into the NW as signified by the grading in Fig. \ref{fig:Model}. 
However, it is assumed that an effective transmission per unit area 
can be assigned to the subsurface region. 
The bulk region is defined as the core atoms for which the transmission 
per unit area approaches the bulk behavior $t_{bulk}$.

The area of the surface region can be expressed as

\begin{equation}
\label{eq:surface}
A_{surf} = \pi [R^2 - (R-\delta)^2] = (2 \pi R) \delta - \pi \delta ^2,
\end{equation}

which for $R \gg \delta$ can be approximated as

\begin{equation}
\label{eq:surface-2}
A_{surf} \approx (2 \pi R)\delta.
\end{equation}

Similarly the area for the subsurface region can be expressed as

\begin{equation}
\label{eq:sub}
A_{sub} = \pi [(R-\delta)^2 -r^2] = 2\pi (R-\delta)d - \pi d^2,
\end{equation}

which for $R \gg d + \delta$ is given as

\begin{equation}
\label{eq:sub-2}
A_{sub} \approx (2 \pi R)d.
\end{equation}

The core region displaying bulk-like behavior has an area given by

\begin{equation}
\label{eq:bulk}
A_{bulk} = \pi r^2 = \pi (R - d - \delta)^2
\end{equation}

which for $R \gg d + \delta$ becomes 

\begin{equation}
\label{eq:surface}
A_{bulk} \approx  \pi R^2.
\end{equation}

The fact that $A_{surf}$ and $A_{sub}$ scale linearly with $R$ whereas 
$A_{bulk}$ scales as $R^2$ for large values of $R$ simply describes 
the fact that the surface and subsurface play decreasing roles as 
the NW's radius is increased. 
An effective transmission per unit area $t$ is then defined for each 
region and the total transmission is expressed as

\begin{equation}
\label{eq:surface}
T = t_{surf}A_{surf} + t_{sub}A_{sub} + t_{bulk}A_{bulk}.
\end{equation}

To explore the role of the surface and subsurface transmission per 
unit area in small cross section NWs we consider four cases.

\textit{Case I: $R \approx \delta$}. 
In this limit, the surface dominates and transport is given purely 
by the surface and represents, for example, the case of an atomic chain. 

\textit{Case II: $R \approx d + \delta$}. 
This condition describes when the cross sectional area of the NW is 
of a scale that a subsurface region is present but the NW diameter 
is not large enough for the core of the NW to display bulk behavior 
for transmission (it should be noted that the subsurface region is not 
sharply defined as represented by the grading of the gray region used 
to define it in Fig. \ref{fig:Model}). 
In the transmission plots of Fig. \ref{fig:1nm-Cross-Section} for the 
unterminated Cu NWs, it can be seen that the subsurface region has a 
lower transmission per unit area compared to the NW surface regions. 
Hence by decreasing the NW diameter from 3 nm to 1 nm, a larger 
transmission per unit area is obtained consistent with the results listed 
in Table \ref{tab:Transmission_Area}. 

\textit{Case III: $r \approx d + \delta $}. 
As the NW radius increases, eventually a core region is formed 
whereby local transmission pathways approach the values obtained in 
bulk copper. 
When this core region is of a comparable dimension to the surface and 
subsurface layer, the transmission per unit area will reflect 
the contributions from the different regions. 
As the subsurface region in unterminated Cu NWs has a lower 
transmission per unit area than the bulk, the overall transmission per 
unit area will increase relative to Case II as the region forms. 

\textit{Case IV: $ R \approx r$}. 
As the NW's radius is further increased, the core region's contribution 
will quickly dominate due to the area increasing as $R^2$ as opposed to the 
linear dependence on NW radius for the surface and subsurface regions. 
Hence the transmission per unit area for the NW will increase with 
increasing $R$ until asymptotically approaching the bulk value. 

The following picture emerges in the case of unterminated NWs.  
For extremely small diameter NWs with diameters of approximately 1 nm, 
surface and subsurface transmissions combine to provide a transmission per 
unit area that is comparable to bulk copper for [100] and [110] 
and less than bulk copper for [111]. 
As the NW diameter increased, for example up to 3 nm, the lower 
subsurface transmission with respect to surface transmission results in a 
net lowering of overall transmission per unit area compared to smaller 
diameter NWs. As the NW diameter is increased further, the 
core region begins to behave like a bulk conductor and the transmission 
rises and asymptotically approaches the bulk transmission per 
unit area. 

The picture needs to be slightly modified for the case of NWs with 
surface oxidation. 
As seen in Fig. \ref{fig:1nm-Cross-Section}, the surface transmission is 
reduced significantly due to surface oxidation. 
From Table \ref{tab:Transmission_Area}, it is seen that the transmission per 
unit area for the 1 nm and 3 nm diameter NWs are comparable. 
Once oxidized, the transmission at the surface is suppressed and the 
transmission of the subsurface region dominates.
We anticipate that the transmission per unit area of the subsurface region 
for a given orientation is similar for 1 nm and 3 nm NWs and this combined with 
the low surface transmission (see Fig. \ref{fig:1nm-Cross-Section}) leads 
to the transmission per unit area for each orientation being approximately 
constant going from 1 nm to 3 nm diameter,  
as reported in Table \ref{tab:Transmission_Area}. 
However, the transmissions of both  1 nm and 3 nm oxidized NWs 
are much lower than 
found for the bulk, and hence larger cross section NWs are required 
before significant bulk-like behavior from the core contributes to the 
net transmission per unit area.  

\section{\label{Conclusion}Conclusion}
The electron transport properties of 1 nm and 3 nm diameter 
copper NWs have been calculated in each case for unterminated and 
surface oxidized NWs of [100], [110] and [111] crystal orientations. 
We find even 1 nm diameter copper NWs to be metallic, however electron 
transmission is strongly dependent on both surface termination and crystal 
orientation. 
Surface oxidation suppresses electron transmission compared to unterminated 
copper NWs,  
consistent with previous reports for copper thin films. 
The [110] oriented NWs consistently show a higher electron transmission 
than [100] oriented NWs which in turn have a higher electron 
transmission than [111] NWs. 
Transmission in unterminated copper NWs is larger at the NW 
surface than below the surface, irrespective of crystal orientation. 
A different picture emerges for surface oxidized NWs wherein 
transmission is lower at the NW surface than in the subsurface for 
[100] and [110] oriented NWs. However while the oxidized [111] NW 
surface transmission is reduced compared to the unterminated NW, 
it nonetheless continues to show larger surface transmission 
than subsurface transmission. 
Transmission per unit area decreases with increasing diameter for 
unterminated NWs but remains approximately constant for the surface 
oxidized NWs studied. 
A simple model of transmission in round NWs which divides the NW 
into regions based on local transmission pathways explains the differing 
behaviors of unterminated and oxidized NWs. 
Briefly, transmission per unit area is lower in the subsurface region than 
at the surface for the unterminated NWs studied, thus when the diameter 
is increased from 1 nm to 3 nm the transmission per unit area of the NW 
decreases as the subsurface region becomes relatively more important. 
In oxidized NWs, the overall transmission per unit area is comparable 
for 1 nm and 3 nm NWs. 
The suppression of the surface transmission in this case increases the 
contribution of the subsurface to overall transmission and a similar 
subsurface transmission per unit area may be expected for 1 nm and 3 nm 
oxidized NWs for the same crystal orientation resulting in similar 
overall transmission per unit area. 
Overall, our results suggest that conductivity in sub 3 nm copper NWs 
is highly sensitive to crystal orientation and 
surface termination, and that careful material preparation and processing will be 
essential in order to maximize conductivity. 

\begin{acknowledgments}
This work was performed as part of the Intel-Tyndall research 
collaboration sponsored by Intel Components Research. 
ASS was funded under an Irish Research Council postgraduate scholarship. 
We are grateful to QuantumWise A/S for providing access to the QuantumWise 
simulation software. 
\end{acknowledgments}

\end{document}